\documentclass{article}

\usepackage{graphicx}
\usepackage{textcomp}
\usepackage{amssymb}
\usepackage{natbib}
\begin{document}
\pagenumbering{arabic}
\title{Estimating the Density of Intermediate Size KBOs from Considerations of Volatile Retention}
\author{Amit Levi and Morris Podolak$^*$ \\
Dept. of Geophysics \& Planetary Science\\
Raymond and Beverly Sackler Faculty of Exact Sciences \\
Tel Aviv University \\
Tel Aviv Israel 69978}

\maketitle 

\noindent
Pages: 36\\
Tables: 0\\
Figures: 5\\

\noindent
$^*$ corresponding author e-mail addresses:  morris@post.tau.ac.il \\

\newpage

\noindent
{\bf Proposed running head:} Estimating KBO densities

\noindent
{\bf Editorial correspondence to:}\\
Prof. Morris Podolak \\
Dept. of Geophysics and Planetary Sciences \\
Tel Aviv University \\
Tel Aviv, Israel 69978 \\
Phone: 972-3-640-8620 \\
Fax: 972-3-640-9282 \\
E-mail: morris@post.tau.ac.il

\newpage

\section*{Abstract}

By using a hydrodynamic atmospheric escape mechanism \citep{amit09} we show how the unusually high mass density of Quaoar could have been predicted (constrained), without any knowledge of a binary companion. We suggest an explanation of the recent spectroscopic observations of Orcus and Charon \citep{delsanti10,cook07}. We present a simple relation between the detection of certain volatile ices and the body mass density and diameter. As a test case we implement the relations on the KBO $2003$ AZ$_{84}$ and give constraints on its mass density.  We also present a method of relating the latitude-dependence of hydrodynamic gas escape to the internal structure of a rapidly rotating body and apply it to Haumea.

\bigskip

\noindent
{\bf Key Words:} ICES; KUIPER BELT; TRANS-NEPTUNIAN OBJECTS

\newpage
 
\section{Introduction}

Jeans escape is often used to estimate the escape rate of volatiles from KBOs \citep[see][for example]{schaller07}.  However, for small bodies hydrodynamic escape may be more important.  A good discussion of atmospheric escape is given in \cite{trafton08}.  In previous work \citep[hereafter paper I]{amit09} we have shown that, with proper modifications, the coronal flow theory, as derived by \cite{parker63}, can be used to describe hydrodynamic atmospheric escape off some KBOs. We found that for an adiabatic hydrodynamic atmospheric escape to occur, the body's diameter $D$ has to be smaller than some maximum value, $D_{max}$, which obeys the relation:
\begin{equation}\label{dmax}
    D_{max}=\sqrt{\frac{3kT_b}{\pi G\rho_bm_g}\frac{\gamma}{\gamma-1}}
\end{equation}    
Here $k$ and $G$ are Boltzmann's and Newton's constants respectively, $T_b$ is the body surface temperature, $\rho_b$ the body mass density, $m_g$ the molecular mass of the atmospheric component of interest, and $\gamma$ its adiabatic index.  

It is important to note that the mechanism of volatile loss and retention for KBOs in hydrodynamic flow is inherently different from that of Jeans escape \citep{schaller07}.  An in-depth discussion of the differences between the two mechanisms can be found in paper I. We simply point out here, that, although the Jeans mechanism may indeed control atmospheric escape from certain large KBOs, it is often a slow process.  Although a low rate, integrated over the age of the solar system can remove the required amount of volatiles, a rapid replenishment by a short-term process such as cryovolcanism will deposit volatiles on at least part of the surface before Jeans escape can fully remove them.  The assumption of Jeans escape must be coupled with the assumption of no fast replenishment otherwise it can lead to erroneous interpretations.  A hydrodynamic escape mechanism removes gas much more quickly, and so avoids this limitation.  A case in point is Enceladus, where jets of water emanate and produce a plume of gas containing CO and other gas species \citep{waite09}.  However CO ice is not found as a surface deposit near the south pole \citep{brown06} because it escapes before it can recondense.        
 
It is commonly assumed that unprocessed water ice in outer solar system objects is amorphous \citep{notesco96,prialnik08,jewitt08}.  However, crystalline water ice was observed on the surfaces of many intermediate size KBOs, e.g. Charon \citep{cook07}, Haumea, 2003 AZ$_{84}$, Orcus, and Quaoar [see \cite{delsanti10} and references cited therein].  The presence of crystalline rather than amorphous ice suggests that the surface must have been heated to a temperature of approximately $100$~K, far above the $30-40$~K surface equilibrium temperatures typical of KBOs. The surface temperature of Pluto, for instance, has been put at about $38$\,K based on surface vapor pressure estimates \citep{delsanti04}.  Although \cite{porter10} have suggested that micrometeorite impact annealing may convert amorphous to crystalline ice in the outer solar system, it is not yet clear if the micrometeorite flux in the Kuiper belt is sufficiently high to process the ice on those bodies.  In any case, the precise heating mechanism is not important.  What is important is that the surface was heated to at least 100\,K for some period of time.  In this paper we consider how these results may be used to estimate the density of KBOs.  In section 2 we present the basic assumptions underlying the model.  In section 3 we consider the relevant timescales.  In section 4 we apply this model to several KBO's for which the required observational data are available.  Here we also extend the model to rapidly rotating bodies.  In section 5 we discuss the results.

\section{Basic Considerations}
Certain volatile species that are mostly frozen at $30-40$~K, will experience fast desorption when the temperature reaches $\approx 100$~K, establishing a transient gaseous atmosphere. If the diameter of a body is larger than $D_{max}$ it will have a strong enough gravitational field to prevent the gas from flowing away hydrodynamically. The gas will then either recondense upon re-cooling of the surface to the equilibrium temperature or flow via tangential winds to colder regions where, again, recondensation would ensue. On the other hand, if the diameter is smaller than $D_{max}$ the volatile atmosphere will escape hydrodynamically. In this case repeated heating of different surface ice patches may eventually clear the surface of the volatiles considered.  If the body allows a particular gas species to flow hydrodynamically into space then that species will not appear in the body surface spectra.  We now ask for which volatiles such a mechanism will apply.  Note that we do not require the loss of large amounts of volatiles.  All we require is that if surface volatiles are vaporized, they be able to leave the body before they recondense on its surface.

For convenience, in what follows, we refer to cryovolcanism as the heating mechanism.  However, we have made no effort to identify the precise mechanism or to estimate its strength.  The fact that the ice was able to undergo the transformation from amorphous to crystalline form, means that some mechanism was able to provide the necessary energy.  Studies by \cite{schmitt89} have shown that the time for amorphous ice to transform is very temperature dependent.  At $T=100$\,K it takes hundreds of years, while at $T=140$\,K it takes on the order of an hour.  If the methane is in the form of a fairly homogeneous layer, then its vapor pressure will be $\sim 4\times 10^4$\,Pa at $T=100$\,K \citep{brownzeig80}.  A surface ice with a vapor pressure given by $P_{vap}(T)$ will sublimate with a flux of \citep{amit09}
$${\cal F}=P_{vap}\sqrt{\frac{m_g}{2\pi kT}}\ \ \ \rm{g\,cm}^{-2}\rm{s}^{-1}$$
If $\rho_{ice}$ is the bulk density of the ice, then the thickness, $\ell$, of the ice layer changes at a rate 
$$\frac{d\ell}{dt}=\frac{{\cal F}}{\rho}$$
The density of solid methane at cryogenic temperatures is of the order of 0.5\,g\,cm$^{-3}$ \citep{bol71}, and this gives a vaporization flux of 7\,g\,cm$^{-2}$\,s$^{-1}$.  If the surface remained at this temperature for the 100\,yr required for amorphous ice to transform to crystalline ice, a layer more than $4\times 10^5$\,km thick would be lost.  At $T=140$\,K, the vapor pressure is $1\times 10^6$\,Pa, and the vaporization flux is some 160\,g\,cm$^{-2}$\,s$^{-1}$.  In the hour it takes the amorphous ice to transform, around 11\,km of ice would be lost.  Any water mixed in with the methane would remain frozen and unless the methane layer was very pure, an ice layer would quickly be formed that would cover any remaining methane.  If the methane is adsorbed on a water ice layer initially, the vapor pressure above the ice would be considerably lower.  We consider this case below.

In what follows we assume that the volatile in question is adsorbed on the water ice.  In general, the bond between a volatile and water ice is stronger than between the volatile and itself, so this assumption gives an upper limit on the escape time.  We further assume that the desorption timescale is the limiting factor for escape.  Because we are considering hydrodynamic flow, once the gas enters the atmosphere and is allowed to flow hydrodynamically, it will escape the body in a very short time (of the order of hours).  We also assume that areas where crystalline ice is observed were heated to at least 100\,K.  If the body is small enough, the vapor will flow adiabatically, as described in paper I, and it will escape hydrodynamically.  The flow will be essentially radial and the gas will not migrate to some colder (shielded) region.  Finally, we argue, on the basis of symmetry, that we expect the KBO surface to be heated more or less uniformly.  It therefore seems likely that if crystalline ice is observed on the surface of such a body, we should not expect large shielded areas where surface volatiles have not been exposed to heating.  However, if such shielded areas existed, and were observed, and if the amorphous ice in these areas were covered by methane or some other material, then our argument would not apply.

\section{Timescales}

Suppose that volatiles are replenished by short-term processes such as cryovolcanism.  If the adsorption timescale is longer than the replenishment timescale, the volatiles cannot escape between replenishments, and volatile loss by the Jeans mechanism is unlikely.  Here we estimate the adsorption timescale. The average time, $\tau$, that a molecule stays adsorbed on an icy surface is given by:
\begin{equation}
    \frac{1}{\tau}=\nu_N e^{\frac{-E_a}{kT}}
\end{equation}        
Where $\nu_N$ is the frequency of oscillation of the adsorbed molecules and $E_a$ is the heat of adsorption \citep{hobbs74}. As in \cite{leger83} we assume for $\nu_N$ the atom surface vibrational frequency which is of the order $10^{14}$~Hz.

CH$_4$ has neither a dipole nor a quadrupole moment, thus it is not expected to interact strongly with OH bonds at the water ice surface. Adsorption is very sensitive to the surface coverage percentage and to the number of favorable adsorbing locations on the icy surface such as defects and amorphous locations. Adsorption measurements of CH$_4$ on water ice I$_h$ result in heats of adsorption, $E_a/k$, in the range $1380-2050$~K, where the lower bound is for ice heated to $238$~K and then re-cooled to $77$~K \citep{chaix97}. The heating probably causes a better ordering of the ice surface and eliminates defects and amorphous structures. The surfaces of KBOs continuously suffer impacts of cosmic rays and solar UV and visible photons \citep{cooper03,cook07} that break water bonds and reduce crystal water ice to amorphous ice. Indeed, observations of crystalline water ice on KBO surfaces are usually accompanied by the detection of some amount of amorphous ice as well. We shall therefore use the upper bound for the heat of adsorption of methane to water ice.

We found no data for the heat of adsorption of acetylene (C$_2$H$_2$), ethylene (C$_2$H$_4$), and ethane (C$_2$H$_6$) on water ice in the literature. However, the heats of sublimation, $\Delta H_{sub}/k$, for the pure volatiles for these hydrocarbons are: $2827$, $2202$ and $2466$~K, respectively \citep{steph87}. The bonding energy of volatiles with water ice is generally stronger than the bonding with their own kind, so the heats of adsorption to water ice will most likely be higher than the values cited. We therefore feel confident that the values for $E_a/k$ for CH$_4$, C$_2$H$_2$, C$_2$H$_4$, and C$_2$H$_6$ fall between $2000-3500$~K. They are unlikely to be much higher than this since no hydrogen bonding exists between these hydrocarbons and water.  For this range of the heat of adsorption to water ice, and for $T_b=35$~K we find $\tau\sim 10^{3}-10^{22}$~yr. Note that the upper bound greatly exceeds the age of the solar system.
 
\cite{cooper03} have shown that the timescale of amorphization of crystalline water ice, along the depth probed in surface spectroscopy, by galactic cosmic rays is of the order of $10^6-10^7$~yr. \cite{cook07} have calculated that the action of solar photons in the range $100-620$~nm is even more effective, with an amorphization timescale of $10^4-10^5$~yr, suggesting a periodic replenishment of crystalline water ice on the surface of Charon, through cryogenic activity, on time scales of less than $10^5$~yr. Therefore, even our lower bound on $\tau$ is only an order of magnitude shorter than the expected time between cryovolcanic eruptions. If some area on the object's surface, perhaps near the poles, reaches temperatures as low as $30$~K then the lower bound for $\tau$ becomes $10^7$~yr. Thus the hydrocarbons we have mentioned above may be considered frozen on the surface (or at least some part of it) between successive cryovolcanic eruptions. When a surface patch of ice is heated to $\approx 100$~K, the average time of adsorption reduces to $10^{-5}-10$~s. In such a case the hydrocarbons considered are promptly released to take an active part in forming a transient atmosphere. The question is whether the body will allow the transient atmosphere to flow away during the excess heating period or not. A clue to the answer is contained in eq.\,\ref{dmax}.

Methanol is particularly interesting in this respect.  Molecular dynamics calculations give a value of $5992$~K for the heat of adsorption of methanol (CH$_3$OH) on water ice.  This high value is due to two hydrogen bonds formed between the two molecular species \citep{picaud06}. Thus even for a temperature of $100$~K the adsorption time would be $\tau\sim 10^4$~yr. Since the re-cooling of cryo-lava takes less than $1$~yr \cite{lorenz96}, the surface will re-cool before methanol can take part in a transient atmosphere. For an ice surface temperature of $35$~K methanol will have an adsorption time of $10^{53}$~yr, much longer than the age of the solar system. \textsl{So methanol should play no part in a transient atmosphere and the body size (i.e. gravity) should play no role in deciding whether methanol shows up in the surface spectroscopy}. It should only be a question of whether some chemical process was able to transform methanol to some other species. 

A tentative corroboration of this argument comes from the possible observation of methanol on the small object \textsl{Bienor} which also, apparently, shows crystalline water ice in its spectrum [see \citep{guilbert09} and references cited therein]. This implies that the  surface was temporarily heated to about $100$~K.  Because of its small diameter of $200$~km, \textsl{Bienor} would require a mass density greater than 41\,g\,cm$^{-3}$ in order to retain methanol. 

CO (and certainly also N$_2$) is highly volatile, having a heat of adsorption to water ice of $1150$~K \citep{Nair70}. Therefore at surface temperatures of $30-40$~K CO will have adsorption times of $\tau\sim 10^{2}-10^{-2}$~s. So even at KBO equilibrium temperatures, in between the periods of excess surface heating, CO (and N$_2$) would be a major atmospheric component that can either escape hydrodynamically or via a fast Jeans mechanism (paper I) and may become depleted in between cryovolcanic eruptions. CO and N$_2$ therefore demand a more rigorous treatment such as that given in paper I, which allows for hydrodynamic escape.

\section{Results}

For each volatile of interest, we need to determine the appropriate values of its polytropic index, $\gamma$. At a temperature of $100$~K the translational and rotational degrees of freedom are active and behave classically whereas molecular vibrational degrees of freedom are largely frozen \citep{loeb}. Using experimental results for the isobaric heat capacity, C$_p$, at $100$~K from \cite{gurvich89} we find $\gamma=$ $1.33$, $1.39$, $1.33$ and $1.30$ for CH$_4$, C$_2$H$_2$, C$_2$H$_4$ and C$_2$H$_6$, respectively. Hydrodynamic atmospheric escape is quite sensitive to the value of $\gamma$.  The freezing out of the vibrational degrees of freedom, which increases the value of the adiabatic index, plays an important part in the theory's ability to match the spectroscopic observations from KBO surfaces. 

[{\bf Figure 1}]

In fig.$1$ we plot the bulk density of four Kuiper belt objects as a function of their diameter.  The error ellipses around each object indicate the uncertainties in the measured masses and radii of these objects.  Also shown is the relation between bulk density and $D_{max}$ (from eq.\,\ref{dmax}) for $T_b=100$\,K for the four hydrocarbons of interest. All objects to the left of a given curve will allow that particular gaseous species to flow hydrodynamically along an adiabat into space. All objects to the right of a given curve have a strong enough gravitational field to confine that particular gas which, upon re-cooling, ought to appear in the object's surface spectrum. The curves for C$_2$H$_6$ and C$_2$H$_4$ fall one on top of the other and are almost indistinguishable. We particularly chose intermediate size KBOs whose size and mass density are known, to a varying degree of accuracy, and can be compared with our theoretical predictions. In addition, the objects selected all show traces of crystalline water ice.  Haumea presents a special case because of its rapid rotation and non-spherical shape, and we discuss it in more detail below.  The error ellipse in this case was calculated using the diameter of a sphere with the same volume, with the uncertainties derived from conditions on an oblate spheroid as given by \citep{rabinowitz06}.

\subsection{Slowly Rotating Bodies}
\begin{figure}[ht]
\centerline{
\includegraphics[scale=0.6]{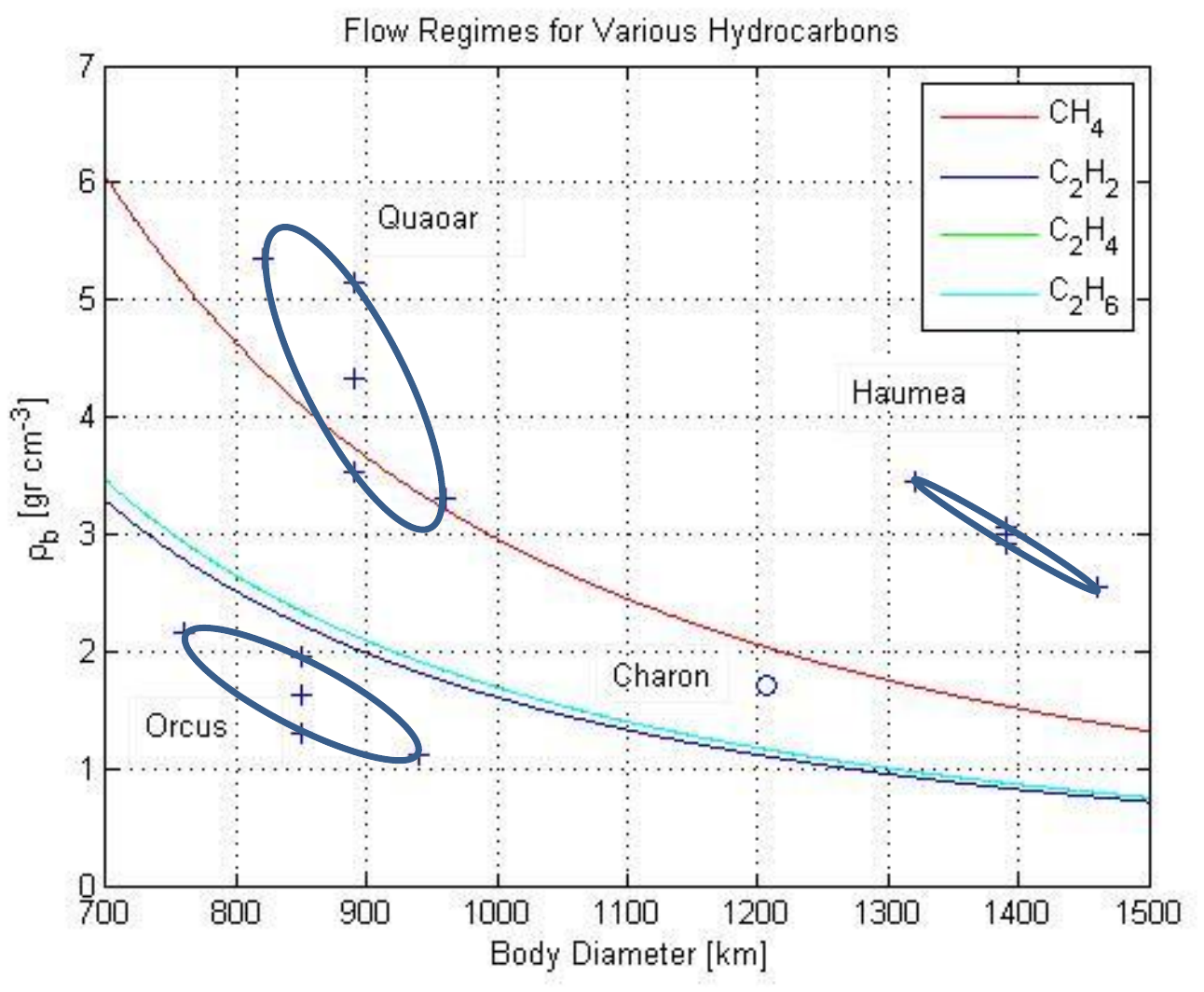}}
\caption{Levi and Podolak, Estimating KBO densities}
\label{fig:flowregime}
\end{figure}

{\bf Orcus}\ -- The diameter and mass are taken from \citep{brnrag10} and  \citep{limetal10}. According to \citep{delsanti10} crystalline water ice is unambiguously detected. As pointed out by these authors, the noise in the data for Orcus is high and Hapke models do not give a unique solution.  Therefore one cannot firmly exclude the presence of ammonia and  methane on the surface. The observed spectrum is best fit by a mixture of water ice (mostly crystalline), ammonium (NH$^+_4$) and ethane. C$_2$H$_2$ and C$_2$H$_4$ were not searched for since ethane was considered more likely, being an irradiation product of methane. It can be seen from fig.\,\ref{fig:flowregime} that, according to our model, Orcus is not expected to retain C$_2$H$_6$ and methane on its surface.  Higher molecular weight hydrocarbons can be present, however, and might explain the weak bands observed in Orcus' spectrum.    

{\bf Quaoar}\ -- Besides crystalline water ice, the presence of methane ice is also confirmed, where addition of ethane improves the quality of the fit to the spectrum [see \citep{guilbert09} and references cited therein]. This is in complete agreement with our model, as shown in fig.\,$1$. Quaoar's diameter is $890\pm 70$~km and with the help of its satellite, Weywot, its density was determined to be $4.2\pm 1.3$~g cm$^{-3}$ \citep{wesley10}. According to eq.\,16, Quaoar must have a mass density of $\rho_b>3.8\pm 0.6$~g cm$^{-3}$ in order to show traces of methane ice on its surface. The uncertainty in this result stems from the uncertainty in Quaoar's reported diameter. We see that the exceptionally high density of this object could have been predicted without any measurement concerning its satellite, Weywot.  

{\bf Charon}\ -- Occultation measurements give Charon's diameter as $1207\pm 3$~km and an average mass density of $1.71\pm 0.08$~g cm$^{-3}$ \citep{sicardy06}. For these values, our model predicts no methane should be visible in the surface spectra, although C$_2$H$_6$, as well as C$_2$H$_2$ and C$_2$H$_4$ could be present. The surface spectrum of Charon was investigated by \cite{cook07} who found evidence for crystalline water ice. Adding methane ice to their model ice mixture did not improve the fit to the observed surface spectra of Charon. We found no mention that C$_2$H$_6$ was searched for.

\subsection{Rapid Rotators}
{\bf Haumea}\ -- Haumea presents a particularly interesting case.  It is a fast rotating elongated body that probably suffered a giant impact \citep{brown07,delsanti10}. \cite{pinilla09} showed that its surface is best described by amorphous and crystalline water ice and that it is probably depleted of volatile ices. As Haumea is a rapid rotator our model does not strictly apply, since we have assumed that the effect of the centrifugal acceleration imparted by the rotating body on the escaping gas is negligible. For Haumea the rapid rotation effectively lowers the depth of the gravitational well. 

Since a rapidly rotating body will deviate from spherical symmetry and there is likely to be a superposition of tangential flow on a rapid radial flow, we need to modify the theory of coronal expansion as developed by E.N. Parker \citep{parker63} for non-radial streamlines by adding the effect of centrifugal acceleration.  The flow velocity, $v$, along the streamline obeys:
$$
v\frac{\partial{v}}{\partial{l}}+\frac{1}{\rho}\frac{\partial{p}}{\partial{l}}+\frac{\partial{\Phi}}{\partial{l}}+\frac{\partial{\Psi}}{\partial{l}}=0
$$
where $l$ is the length along the streamline, $\rho$ and $p$ are the gas mass density and pressure respectively, $\Phi$ and $\Psi$ are the gravitational and centrifugal potentials external to the body. Assuming a polytropic relation between the gas pressure and density, with a polytropic index $\gamma$, one obtains:
\begin{equation}\label{stream}
\frac{\partial}{\partial l}\left (\frac{1}{2}v^2+\frac{\gamma}{\gamma-1}\frac{p_0}{\rho_0^\gamma}\rho^{\gamma-1}+\Phi +\Psi \right)=0
\end{equation}
The subscript zero denotes the surface value of the given parameter.  The expression in parenthesis will be constant along the streamline. 

As noted by \cite{parker63}, conservation of matter along the streamline may be expressed as:
$$
\rho(l)v(l)A(l)=\rho_0v_0A_0
$$
where $A$ is the cross section of the flow tube. Combining this with eq.\,\ref{stream} and defining the relative cross section $F\equiv A/A_0$ gives: 
\begin{equation}\label{coronal}
\frac{1}{2}v^2(l)+\frac{\gamma}{\gamma-1}\frac{p_0}{\rho_0}\left(\frac{v_0}{vF}\right)^{\gamma-1}+\Phi(l)+\Psi(l) =\frac{1}{2}v_0^2+\frac{\gamma}{\gamma-1}\frac{p_0}{\rho_0}+\Phi_0+\Psi_0
\end{equation}
This is Parker's coronal flow equation including the centrifugal potential. 

If the streamlines are indeed non-radial then $l$, the distance along the streamline, cannot be approximated by the radial coordinate $r$.  A more natural choice for the independent spatial coordinate is the gravitational potential itself.  Following Parker, and including some changes due to the introduction of the centrifugal acceleration, we define the following dimensionless parameters:
$$\varepsilon\equiv\frac{v}{v_0}\chi^{\frac{1}{\gamma+1}}$$
$$\chi\equiv\frac{\gamma-1}{2\gamma}\frac{\rho_0v^2_0}{p_0}$$ 
$$\phi\equiv-2\frac{\Phi}{v^2_0}\chi^{\frac{2}{\gamma+1}}$$
$$\psi\equiv-2\frac{\Psi}{v^2_0}\chi^{\frac{2}{\gamma+1}}$$
With the help of these dimensionless variables eq.\,\ref{coronal} can be rewritten as
\begin{equation}
\varepsilon^2(\phi)+\left[F(\phi)\varepsilon(\phi)\right]^{1-\gamma}-\phi-\psi(\phi)=\chi^{\frac{2}{\gamma+1}}+\chi^{\frac{1-\gamma}{\gamma+1}}-\phi_0-\psi_0
\end{equation}

Let us examine the asymptotic behavior of this last equation along the streamline as $\phi\rightarrow 0$. It is reasonable to assume that far from the body streamlines diverge, causing $F$ to diverge.  In addition, $\psi$ will vanish. After some algebra \cite[see][]{amit09} we again end up with two branches for the asymptotic solution far from the body, an upper branch and a lower branch, given to zero order by:
$$
\varepsilon_{upper}\approx\sqrt{\chi^{\frac{2}{\gamma+1}}+\chi^{\frac{1-\gamma}{\gamma+1}}-\phi_0-\psi_0}
$$ 
\begin{equation}
\varepsilon_{lower}\approx\left( \chi^{\frac{2}{\gamma+1}}+\chi^{\frac{1-\gamma}{\gamma+1}}-\phi_0-\psi_0\right)^{\frac{1}{1-\gamma}}F^{-1}
\end{equation}
These are the same as those given by Parker, but with the addition of the $\psi_0$ term. Several complexities are introduced due to the streamlines' non-radial form that have to do with the way the flow transforms from subsonic to supersonic flow. Nonetheless it is clear from the asymptotes that the establishment of a steady-state solution far from the body requires:
\begin{equation}
\chi^{\frac{2}{\gamma+1}}+\chi^{\frac{1-\gamma}{\gamma+1}}-\phi_0-\psi_0\geq 0
\end{equation}
Reverting back to our original notation this may be written as:
\begin{equation}
v^{\frac{4}{\gamma+1}}_0+2v^{\frac{2(1-\gamma)}{\gamma+1}}_0\left[\frac{\gamma}{\gamma-1}\frac{p_0}{\rho_0}+\Phi_0+\Psi_0\right]\geq 0
\end{equation}
In the limiting case where the body is massive enough to prevent the streamline from forming (i.e. $v_0\rightarrow 0$) the inequality turns into an equality and since $\gamma>1$ this reduces to the condition:
\begin{equation}\label{stedst}
\frac{\gamma}{\gamma-1}\frac{kT_0}{m_g}+\Phi_0+\Psi_0=0
\end{equation}

In cases where the centrifugal acceleration is of importance the object rotates quickly enough so that it deviates (perhaps considerably) from spherical symmetry.  Let us assume the body behaves as a liquid rotator in hydrostatic equilibrium resulting in a problem independent of longitude. The shape of an ellipsoid of revolution may be formulated as \citep [see, e.g.][]{zhartru78}
\begin{equation}
r_0^2(\theta)=a^2\left[ \frac{\cos^2\theta}{(1-e)^2}+\sin^2\theta\right] ^{-1}
\end{equation}
Here $r_0$ is the radial coordinate at the body's surface, $\theta$ is the colatitude, $a$ is the equatorial radius, and $e$ is the flattening parameter that, given the polar radius $b$, is defined by 
$$
e\equiv\frac{a-b}{a}
$$
The gravitational potential on the surface of the ellipsoid may be expanded using gravitational moments as follows:
\begin{equation}
\Phi_0=-\frac{GM}{r_0(\theta)}\left[1-\sum_{n=1}^\infty\left(\frac{a}{r_0(\theta)}\right)^{2n}J_{2n}P_{2n}(\cos\theta) \right]
\end{equation}
where $G$ is the universal gravitational constant, $M$ is the body's mass, $J_n$ are the gravitational moments and $P_n$ are Legendre polynomials.  The centrifugal potential on the body's surface will be
\begin{equation}
\Psi_0=\frac{1}{2}\Omega^2_0r^2_0\sin^2\theta 
\end{equation}
where $\Omega_0$ is the object's angular velocity.  Inserting all of this into eq.\,\ref{stedst} results in the following restriction on the flow:
$$
\frac{\gamma}{\gamma-1}\frac{kT_0}{m_g}-\frac{GM}{a}\sqrt{\frac{\cos^2\theta}{(1-e)^2}+\sin^2\theta}
\left[1-\sum_{n=1}^\infty\left(\frac{\cos^2\theta}{(1-e)^2}+\sin^2\theta\right)^{n}J_{2n}P_{2n}(\cos\theta) \right]+
$$
\begin{equation}\label{rotrest}
+\frac{1}{2}\Omega_0^2a^2\sin^2\theta \left[\frac{\cos^2\theta}{(1-e)^2}+\sin^2\theta\right]^{-1} =0
\end{equation}
In the case where the body does not rotate eq.\,\ref{rotrest} reduces to the restriction given in \cite{amit09}.

We now consider the case of Haumea and a possible methane atmosphere.  It is not clear whether Haumea is an elongated ellipsoid or a flattened spheroid \citep[see][]{rabinowitz06}.  \cite{ragozzine09} have adopted a tri-axial model.  Nonetheless, we shall assume that the body can be approximated by a Maclaurin spheroid and apply eq.\,\ref{rotrest}.  While this may not be strictly correct, it is useful in exploring some surprising consequences of rapid rotation for atmospheric escape.  The body parameters are taken from \cite{rabinowitz06}.  We take a mass of $M=4.21\times 10^{24}$~g, an average mass density of $\bar{\rho}\approx 2.6$\,g\,cm$^{-3}$, a polar radius of $\sim 500$\,km, and an equatorial radius of $\sim 800$\,km. The rotation period is taken to be $3.9154$\,hr.  The quadrupole moment, $J_2$, is not known. \cite{ragozzine09} argued that $J_2$ ought be large due to Haumea's high amplitude light curve and adopted a value of $0.244$ for their tri-axial model. 

\begin{figure}[ht]
\centerline{
\includegraphics[scale=0.6]{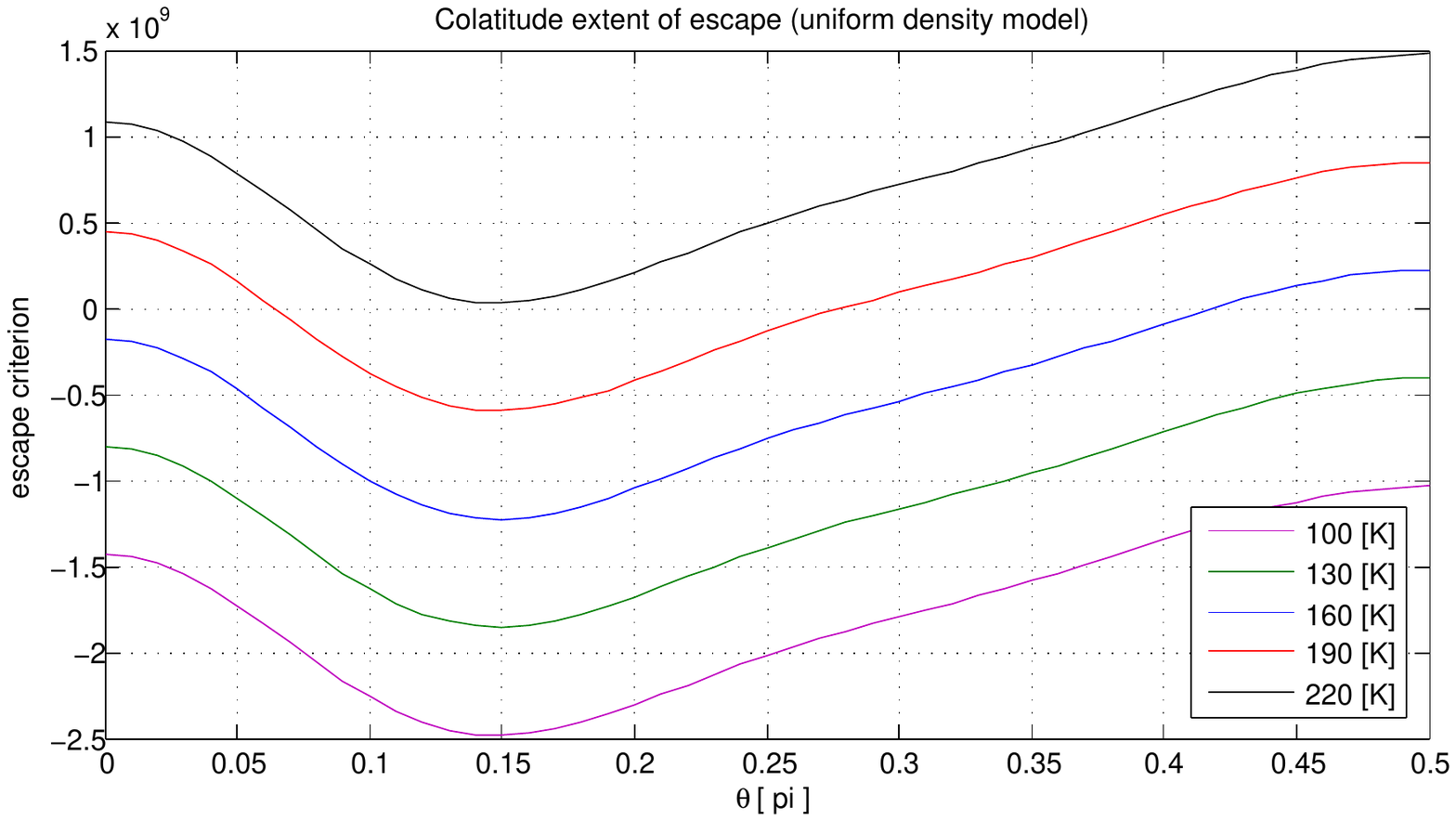}}
\caption{Levi and Podolak, Estimating KBO densities}
\label{fig:constantdensity}
\end{figure}

In order to understand the co-latitudinal restriction on the flow we examine a simple scenario; that of a constant density body.  We define a rotational parameter, $q$, as:
\begin{equation}\label{rotpar}
q\equiv \frac{3\Omega^2_0}{4\pi G\bar{\rho}}=0.2723
\end{equation}
According to \cite{zhartru78} we have for the constant density model:
$$
e=\frac{5}{4}q\left(1+\frac{15}{56}q+\frac{925}{1568}q^2\right)\approx 0.3801
$$
$$
a=\frac{b}{1-e}\approx 806~[km]
$$
$$
J_2=\frac{1}{2}q\left(1-\frac{5}{14}q+\frac{25}{98}q^2\right)\approx 0.1255
$$
$$
J_4=-\frac{15}{28}q^2\left(1-\frac{5}{7}q\right)\approx -0.032
$$
\begin{equation}
J_6=\frac{125}{168}q^3\approx 0.015
\end{equation}
In fig.~\ref{fig:constantdensity} we plot the LHS of eq.\,\ref{rotrest}, the {\it escape criterion}, from the north pole ($\theta=0$) to the equator ($\theta=\pi/2$) for a body with uniform density. Adiabatic flow is allowed only for colatitudes for which the escape criterion is positive. Thus, for the conditions we have assumed, adiabatic flow cannot be supported at a surface temperature of $130$\,K. However, if the surface could be heated to $160$\,K, adiabatic streamlines may form for $\theta > 0.42\pi$, corresponding to latitudes in the range $[-15^\circ\  15^\circ]$.  Surface temperatures higher than about $220$~K are required to support a global adiabatic escape.

\begin{figure}[ht]
\centerline{
\includegraphics[scale=0.6]{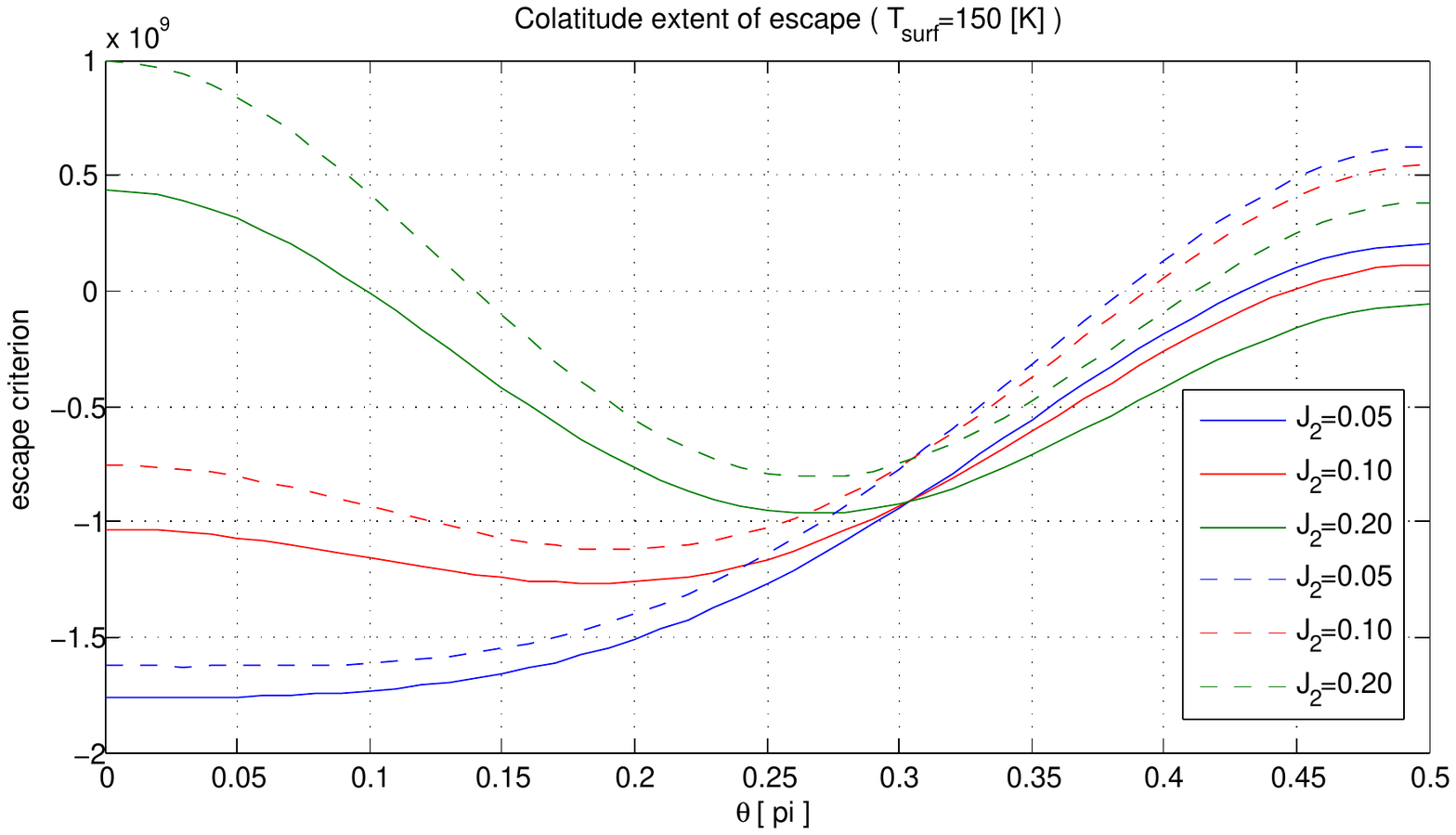}}
\caption{Levi and Podolak, Estimating KBO densities}
\label{fig:escapebehaviour}
\end{figure}

Taking the body's mass and volume as constant while varying $J_2$ has the effect of changing the body's internal density distribution.  The effect of such a variation on the escape criterion is shown in fig.~\ref{fig:escapebehaviour}, assuming a surface temperature of $150$~K. The solid curves show the value of the escape criterion for three cases: $J_2=0.05$, 0.1, and 0.2. Since the object's mass is derived independently of its shape we also consider a $10\%$ increase in the equatorial radius (from $800$~km to $880$~km) while keeping the same mass. This is shown for the same three choices of $J_2$ by the dashed curves.  For all the cases presented in the figure adiabatic streamlines cannot be supported in mid-latitudes.  For values of the quadrupole moment lower than the value corresponding to uniform density ($\approx 0.1$), adiabatic streamlines can be supported only in a region around the equator. for the case of $J_2=0.1$ the latitudinal range allowing for adiabatic gas escape will increase from $[-10^\circ\ 10^\circ]$ to $[-20^\circ\ 20^\circ]$ when the equatorial radius is increased by $10\%$. This is due to the increase in centrifugal acceleration.  

\begin{figure}[ht]
\centerline{
\includegraphics[scale=0.6]{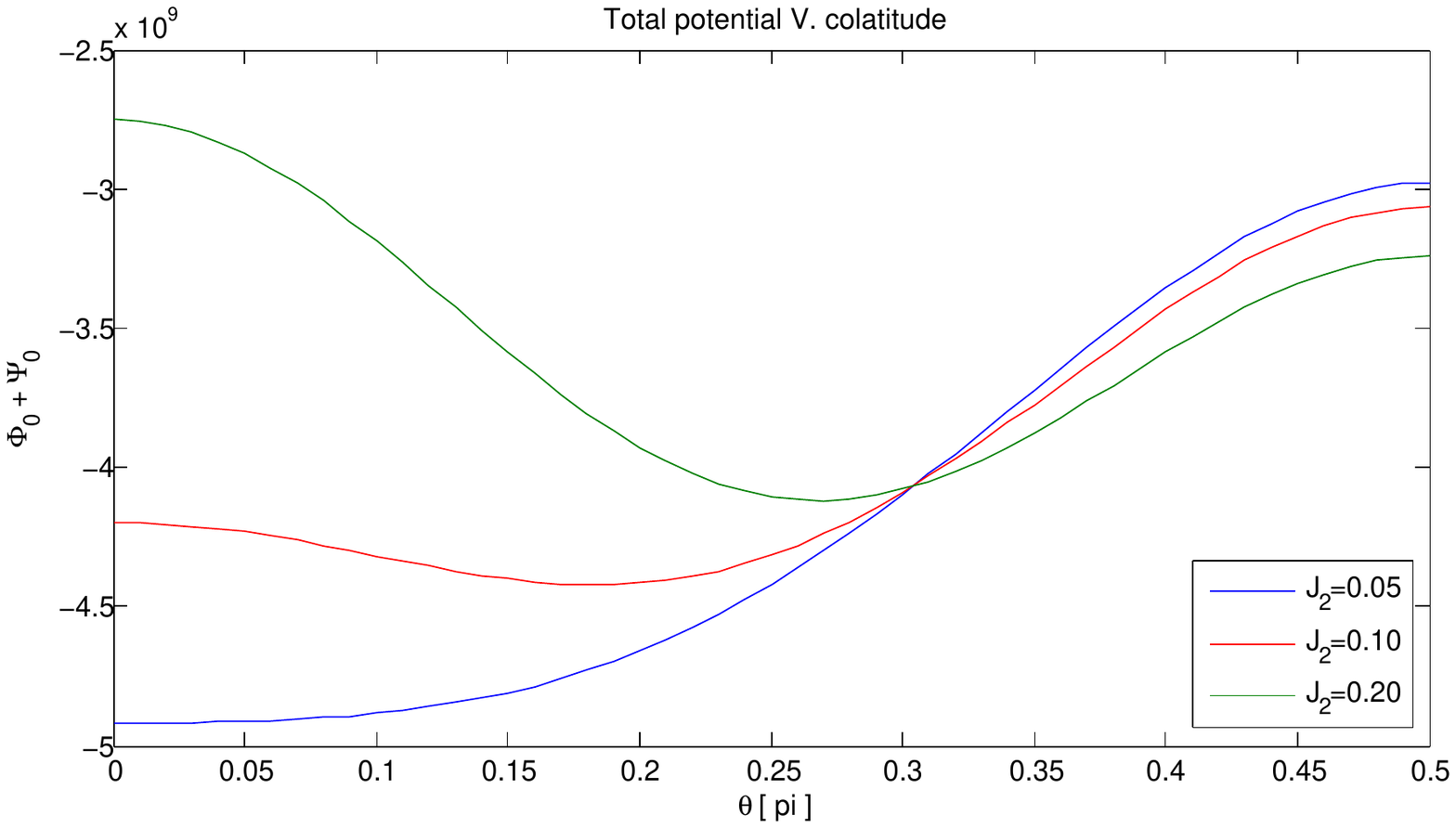}}
\caption{Levi and Podolak, Estimating KBO densities}
\label{fig:potentialcolatitude}
\end{figure}

For $J_2=0.2$ adiabatic streamlines can be supported from the polar region, while only for the case of the larger equatorial radius there will be equatorial flow as well.  This surprising behavior may be explained as follows: A low enough value of $J_2$ represents a differentiated body with a dense core. Since the equatorial dimension is larger than the polar dimension, an observer at the pole will be closer to the core, and feel a stronger attraction from it, while the gravitational attraction from the lobes of the oblate mantle will cancel to some extent due to symmetry. Therefore, a denser core will result in a higher gravitational attraction at the poles and gas escape will be confined to the equator. Increasing the quadrupole moment means that more of the dense material is in the object's envelope rather than the core.  This decreases the gravitational attraction at the pole while somewhat increasing the gravitational attraction at the equator. As a result polar atmospheric escape is easier. In contrast to the poles, no cancellation due to symmetry occurs at mid-latitudes so the gravitational attraction there will remain high.  This can be seen in fig.~\ref{fig:potentialcolatitude}, where we have plotted the total (gravitational + centrifugal) potential as a function of co-latitude for the same three choices of $J_2$.  The behavior we describe is clearly seen.

\begin{figure}[ht]
\centerline{
\includegraphics[scale=0.6]{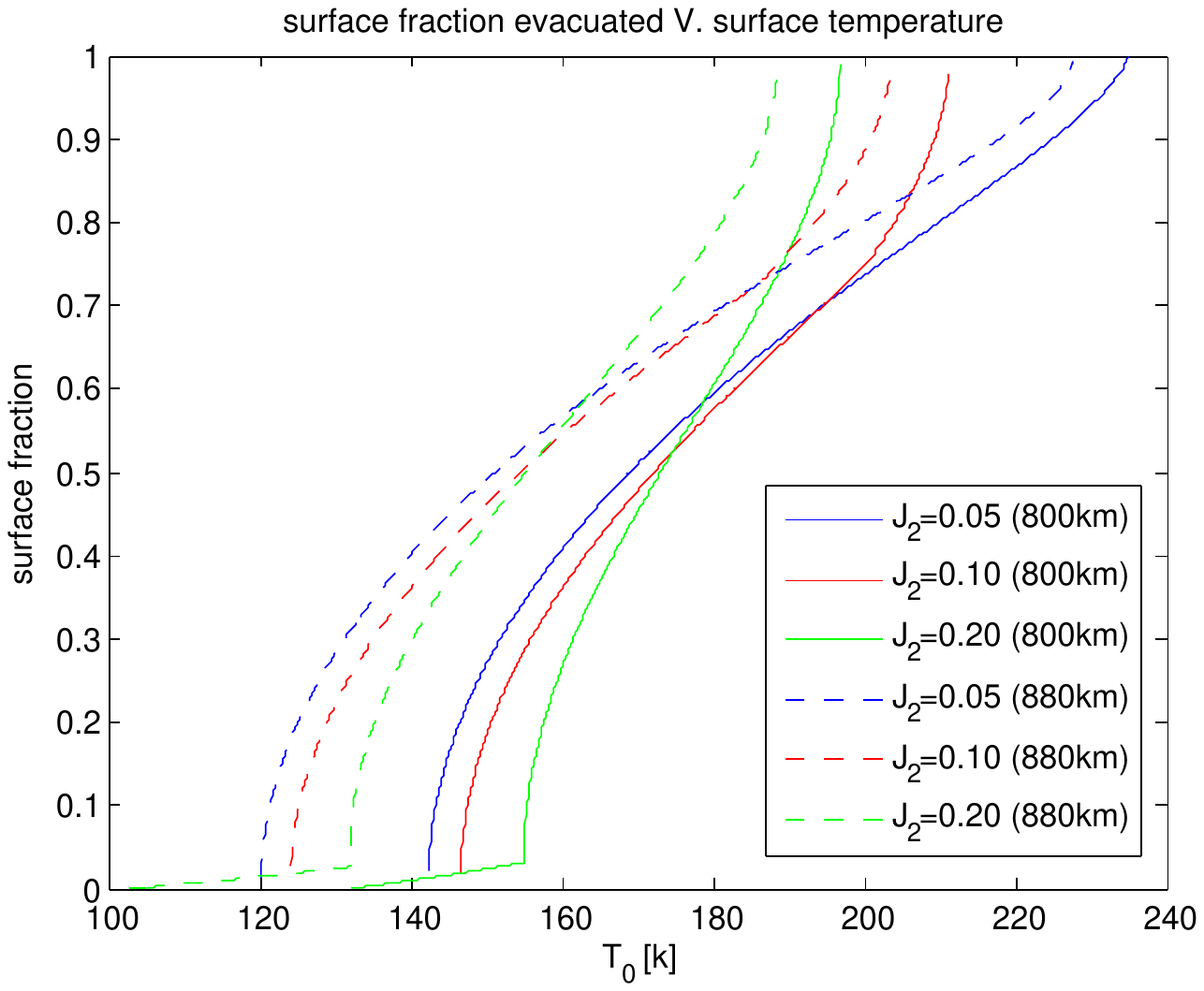}}
\caption{Levi and Podolak, Estimating KBO densities}
\label{fig:surffraction}
\end{figure}

As methane sublimates from a given area it will flow (restricted by the escape criterion) along a streamline into space, clearing that region of methane ice. In fig.~\ref{fig:surffraction} we present the surface fraction evacuated as a function of surface temperature, for the three gravitational quadrupole moments and two equatorial radii considered in fig.~\ref{fig:escapebehaviour}. Due the object's geometry, a given latitudinal range near the equator corresponds to a larger surface fraction than would the same range near the pole. 
For the case of $J_2=0.05$ the flow begins near the equator at low temperature and approaches polar regions monotonically for higher surface temperatures. This results in a rapid increase in surface fraction evacuated. Similar behavior is seen for $J_2=0.1$ up to a certain high surface temperature where polar escape begins as well making the surface fraction evacuated increase even more rapidly. For $J_2=0.2$ flow begins from polar regions resulting in a rather slow increase of surface fraction evacuated per increase of surface temperature up to the temperature for which equatorial escape begins to take effect.      

From fig.~\ref{fig:surffraction} it is seen that at surface temperatures about half to two thirds of those required for global atmospheric escape, streamlines evacuating sublimating gases may already emanate from a large fraction (tens of percent) of the body's surface.  If the axis of Haumea's rotation is perpendicular to our line of sight as suggested by \cite{ragozzine09} and \cite{rabinowitz06} then the equatorial region carries more weight observationally.  In such a scenario the value we estimated for the surface fraction evacuated will be a lower limit to the value observed from Earth.  For $a=800$~km a high depletion of surface methane would suggest a differentiated body since surface temperatures, even during extreme heating, do not go above $150$~K. For the case of $a=880$~km the same would be true if the surface temperatures do not go above $130$~K.

\section{Discussion}
As we have shown, the surface spectra of the four intermediate-size KBOs discussed can be understood in the context of surface heating followed by rapid, hydrodynamic escape. The major advantage of this model is that very simple relations can be obtained between detection of certain volatiles and the body's diameter and density. We suggest the following scheme for bodies that show evidence of crystalline water ice.

Given a body with a measured diameter $D\pm\Delta D$, that shows traces of CH$_4$ ice on its surface, its average mass density will obey:
\begin{equation}
    \rho_b>2.98\left[\frac{1000\,{\rm km}}{D}\right]^2\left[1\pm 2\frac{\Delta D}{D}\right]\,{\rm g\,cm}^{-3}
\end{equation}
Since the curves for C$_2$H$_6$, C$_2$H$_2$ and C$_2$H$_4$ are almost indistinguishable, as seen in fig.\,1, 
we suggest that, if either one of these three hydrocarbons is observed \textsl{while no methane ice is observed} then:
\begin{equation}
   1.71\left[\frac{1000\,{\rm km}}{D}\right]^2 \left[1\pm 2\frac{\Delta D}{D}\right] <\rho_b<2.98\left[\frac{1000\,{\rm km}}{D}\right]^2 \left[1\pm 2\frac{\Delta D}{D}\right] \,{\rm g\,cm}^{-3}
\end{equation} 
In case none of the four hydrocarbons (CH$_4$, C$_2$H$_2$, C$_2$H$_4$ and C$_2$H$_6$) is observed we expect:
\begin{equation}
    \rho_b<1.71\left[\frac{1000\,{\rm km}}{D}\right]^2 \left[1\pm 2\frac{\Delta D}{D}\right] \,{\rm g\,cm}^{-3}
\end{equation}
 
As an interesting test case, let us consider the object $2003$ AZ$_{84}$. Crystalline water ice was detected on its surface \citep{guilbert09}. Some discrepancy exists concerning its diameter. \cite{stansberry08} give a value of $686\pm 95$~km whereas \cite{muller10} report a value of $896\pm 55$~km. If methane ice will be discovered on its surface then, according to our model, its density would have to be higher than $6.3\pm 1.7$ or $3.7\pm 0.5$\,g cm$^{-3}$, depending on its diameter. If CH$_4$ ice is not detected but either C$_2$H$_2$, C$_2$H$_4$, or C$_2$H$_6$ are observed, the minimum density must be either $3.6\pm 1.0$ or $2.1\pm 0.3$\,g cm$^{-3}$, respectively. If none of the four hydrocarbons is seen on the surface then the latter values are upper limits to the density. If the larger (smaller) estimate for the diameter is confirmed and methane (ethane) ice is detected then our model predicts this body has a Quaoar-like mass density, implying another KBO is asteroid-like, covered by a thin sheet of ice. This would have important implications for the way we view KBOs.  If the body is rapidly rotating and has a large oblateness, as is the case with Haumea, observations of the surface coverage of volatile ice can be used to estimate the amount of differentiation the body has undergone. 

\section*{Acknowledgement}
We wish to thank an anonymous referee for helpful comments.

\bibliographystyle{icarus2}
\bibliography{amit3} 

\newpage
\section*{Figure Captions} 
Figure 1 - Atmospheric escape regimes for four hydrocarbon species compared with four KBOs. Error ellipses indicating the most likely uncertainties in the observed parameters are given.  Being to the left (right) of a given curve means the gaseous species can (cannot) escape hydrodynamically.  Haumea is a special case because of its non-spherical shape and rapid rotation, and is discussed separately in the text.\\
Figure 2 - Escape criterion (LHS of eq.\,\ref{rotrest}) as a function of co-latitude for different values of the surface temperature.  Hydrodynamic escape occurs only when the escape criterion is positive (see text).\\
Figure 3 - Escape criterion (LHS of eq.\,\ref{rotrest}) as a function of co-latitude for different values of $J_2$ and a surface temperature of 150\,K.  Solid curves are for an assumed equatorial radius of 800\,km and dashed curves are for an assumed radius of 880\,km.\\
Figure 4 - Total potential (gravitational plus centrifugal) on the surface of an ellipsoid of revolution as a function of co-latitude for different values of $J_2$.\\
Figure 5 - Fraction of the body's surface from which adiabatic streamlines can be energetically supported as a function of surface temperature for different assumed values of $J_2$.  The solid curves are for an assumed equatorial radius of 800\,km.  The dashed curves are for a radius of 880\,km.
 
\end{document}